# Fundamental limits of accuracy and precision in single-molecule biosensors

Tuhin Chakrabortty[*], Manoj M Varma[*#]

*Center for Nano Science and Engineering, Indian Institute of Science, Bangalore, India
# Robert Bosch Center for Cyber-Physical Systems, Indian Institute of Science, Bangalore India

**Abstract:**

**Physical limit of molecular sensing has been extensively studied in biological systems. Biosensors are engineered equivalents of molecular sensors in living systems and play a critical role in disease diagnosis and management. Investigation into the physical limits of biosensors could have major beneficial impact on early disease diagnosis. Here, we present an extension of the classical works on molecular sensing limits of living systems to the realm of biosensors. Two approaches are proposed to estimate concentration with noisy biosensors. We find a trade-off between precision and accuracy.**

Precise molecular sensing is crucial for living systems. Experimental evidence suggests that single bacteria such as *E. coli* can detect chemical concentrations of the order of a few nanomolar [1]. Similarly, amoebas have been shown to be capable of responding to concentration differences as low as 10 molecules across the cell [2, 3]. There has been an active interest in theoretically understanding the mechanisms and accuracy of the systems since the pioneering work by Berg and Purcell [4]. Berg and Purcell calculated the precision of concentration sensing by a single receptor. They considered a scenario in which the cell estimates the ligand concentration by monitoring the occupancy of a receptor to which the ligand molecules bind and unbind. Specifically, for a single receptor, the cell only receives the binary time-series signal of a finite duration $T$ which consists of bound and unbound time intervals from which it needs to estimate the concentration of the external ligand molecule (Figure 1a, 1c). Berg and Purcell showed that a cell can achieve this by calculating the fraction of time the receptor spends being bound with the ligand molecules. The bound fraction of time is a function of the concentration of the ligand molecules. By considering the time correlation of the ligands bound to the receptors, they analytically calculated the variance of the estimated concentration which defines the fundamental limit of precision. Although recent works have shown that the Berg and Purcell limit (BP limit) can be broken by energy-consuming multi-state downstream signaling networks, for equilibrium systems, the BP limit remains valid [5, 6, 7, 8].

Biosensing devices are required for critical applications such as early detection of cancer which significantly improves the survival rates of the patients [9]. Therefore, investigations on the physical limits of molecular sensing in the context of biosensors is an important direction to pursue. Single-molecule sensors such as those based on nanopores almost perfectly map to the well-studied two-state model of a single receptor system. Nanopore sensors consist of a thin sheet of a solid material such as Silicon Nitride, with a 1-10 nanometers diameter pore separating two fluid reservoirs [10]. The nanopore device is immersed in an electrolyte where application of an electric field drives an ionic current through the pore. When a molecule of interest passes through the pore, it blocks the ionic current (Fig 1d). The time-series signal of ionic current is analogous to the binding-unbinding signal due to the ligand-receptor interactions in the cellular context and such a time-series can be used to infer the concentration of the target molecule (Fig 1b) [11]. Therefore, one would expect the analyses of Berg and Purcell and subsequent works on cellular sensing to be applicable for these systems as well. However, while there is a strong analogy between the two systems, there are two key aspects that are different compared to the previous studies on cell sensing, namely, a) the idea of accuracy of sensing in contrast to the precision of sensing, and b) the influence of measurement noise.

In the context of cellular sensing, Berg and Purcell and some subsequent works considered two sources of external noise which corrupt the measurements of the receptors, a) stochastic transport of the ligand molecules to the receptor via diffusion, and b) the binding and unbinding of the ligand molecules to the receptor after they have arrived at its surface [4, 12, 13, 6]. Some recent reports have also considered the effects of the internal noise in the form of noisy cascades of downstream signaling processes [14, 15]. However, the effect of an independent additive noise does not appear to be treated in detail in the existing literature on cell-sensing limits. Measurement noise is an unavoidable part of biosensors. In particular, the measurement noise in many biosensing systems is colored and often there is a strong $1/f$ dependence of the measured noise spectrum, including in nanopore sensors [16].

In this letter, we revisit the Berg and Purcell limit in the context of a biosensor, specifically, single-molecule biosensors. To this end, we consider a single receptor with independent additive measurement noise, exposed to ligand molecules at concentration $c_0$. To obtain an estimate, $\hat{c}$, of the ligand molecules from the binding-unbinding time-series signal, we consider two approaches, namely, a) Direct Averaging (DA) in which we directly average the time-series signal and b) Threshold-based Averaging (TA) where we average the binary signal after thresholding the original noisy signal. Threshold-based determination of transitions from 'open' and 'closed' states of a nanopore from the raw (noisy) current traces is commonly employed [17] justifying our approach. The typical metric considered in Berg-Purcell and subsequent analyses is the relative variance, $\delta\hat{c}^2/\langle\hat{c}\rangle^2 = \langle(\hat{c}-\langle\hat{c}\rangle)^2\rangle/\langle\hat{c}\rangle^2$, which represents the precision of sensing, i.e. the relative error on the estimate. The possibility of a bias in the estimate, i.e. a deviation between the average value of the estimate $\langle\hat{c}\rangle$ and the true concentration $c_0$ has not been considered in previous analysis because direct averaging, as considered in the original Berg and Purcell analysis, provides an unbiased estimate of the concentration even in the presence of noise. In contrast, the threshold-based approach leads to a biased estimate of the concentration. Interestingly, the estimation error of the threshold-based approach can surpass the Berg and Purcell's limit leading to a bias-accuracy trade-off. To address the estimation bias in the threshold-based approach, we define a new metric called relative deviation, defined as $\langle\hat{c}-c_0\rangle^2/c_0^2$. The relative deviation is a measure of the accuracy of sensing, i.e. how close is the estimate to the true concentration. The relative deviation and the relative variance will coincide for unbiased estimates.

We use both analytical derivation and Monte-Carlo simulations to compare the direct averaging and threshold-based approaches. Although not conveniently tractable analytically, unlike Gaussian white noise, given the relevance of $1/f$ noise in real systems, we explore its effect numerically. We see that both types of measurement noise have similar qualitative effects.

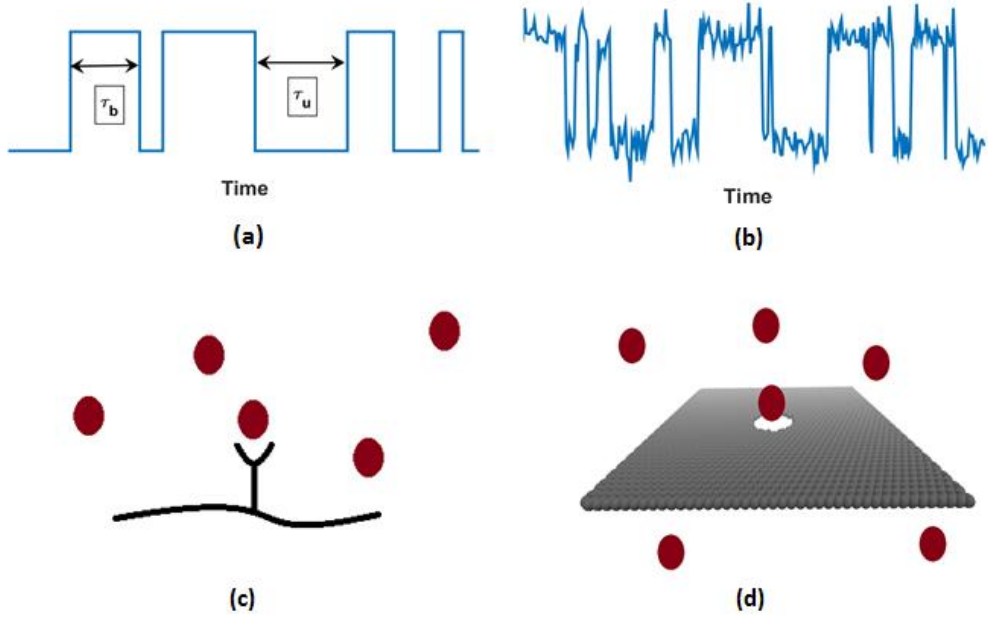

Figure 1 (a) Schematic diagram of the binary signal model in Berg and Purcell, (b) Schematic representation of a noisy nanopore signal (c) Schematic diagram of the interaction of a cell surface receptor and environmental ligand molecules (shown as red circles) (d) Interaction of an artificially engineered nanopore sensor and ligand molecules present in the environment

**Results**

We consider a single receptor sensor in a noisy environment that represents a single nano-pore system described in Figure1(d). We define the signal $m(t)$ observed by the sensor as

$$m(t) = B(t) + \xi(t) \qquad (1)$$

where $B(t)$ is the binary signal which represents the transitions from $0 \to 1$ and $1 \to 0$. Similar to [4], the transition probabilities for $0 \to 1$ and $1 \to 0$ for $B(t)$ are proportional to $k_+c$ and $k_-$ respectively, where $c$ is the concentration of the ligand molecules and $k_+$ and $k_-$ are the binding and unbinding rates of the ligand molecules with the receptor respectively. The noise term ($\xi(t)$) is independent of the binary signal $B(t)$. We consider two cases 1) $\xi(t)$ is a zero mean Gaussian white noise, defined as $\langle \xi(t) \rangle = 0$, and $\langle \xi(t' + \tau_1)\xi(t' + \tau_2) \rangle = \sigma_\xi^2 \delta_{\tau_1 \tau_2}$, where $\delta_{\tau_1 \tau_2}$ is the Kronecker delta function. 2) $\xi(t)$ is a zero mean Gaussian $1/f$ noise, defined as $\langle \xi(t) \rangle = 0$, $\langle \xi(t)^2 \rangle = \sigma_\xi^2$ and $S_\xi(f) = k/f$, where $S_\xi(f)$ is the power spectral density of the noise, $f$ is the frequency and $k$ is an arbitrary constant.

We first take the approach of direct averaging (DA) of the measured signal $m(t)$. Since $\xi(t)$ is zero mean, directly averaging the signal $m(t)$, we get the exact result as Berg and Purcell from which the estimated concentration $\hat{c}_{DA}$ can be calculated as

$$\hat{c}_{DA} = \frac{m_T}{1-m_T}\left(\frac{k_-}{k_+}\right) \tag{2}$$

where $m_T$ is the average of the measured signal $m(t)$, $k_+$ and $k_-$ are the binding and unbinding rates respectively. However, due to the presence of the noise, the variance of the estimated concentration increases. For the case of Gaussian white noise, we can calculate the relative variance of $\hat{c}_{DA}$ as [See Appendix 1 for details]

$$\left(\frac{\delta \hat{c}_{DA}}{\hat{c}_{DA}}\right)^2 = \frac{1}{Tm_T}\left(2\tau_b + \frac{\sigma_\xi^2}{m_T(1-m_T)^2}\right) \tag{3}$$

where $T$ is the integration time and $\tau_b$ is the average time the receptor spends bound to the ligand. The first term of Equation (3) is the same as the BP limit and the second term is due to the measurement noise. Expectedly, Equation (3) reduces to BP limit for $\sigma_\xi = 0$. The relative variance decreases monotonically with increasing measurement time (Figure 2a). The relative deviation, which as we discussed represents the accuracy of sensing, is defined as $\frac{\langle \hat{c}_{DA} - c_0 \rangle^2}{c_0^2}$. As the estimate is unbiased, $\langle \hat{c}_{DA} \rangle = c_0$ making the relative deviation coincide with the relative variance.

However, this picture changes when we consider the threshold-based averaging (TA) approach. Here, prior to averaging, the measured signal $m(t)$ is thresholded to obtain a binary signal $\tilde{m}(t)$. $\tilde{m}(t)$ can be thought of as the original binary signal $B(t)$ corrupted by spurious events due to noise. Since $\tilde{m}(t)$ is a binary signal, if $\tilde{k}_+$ and $\tilde{k}_-$ are the modified binding and unbinding rates, then following Berg and Purcell the concentration can be estimated from $\tilde{m}(t)$ as

$$\hat{c}_{TA} = \frac{\tilde{m}_T}{1-\tilde{m}_T}\left(\frac{\tilde{k}_-}{\tilde{k}_+}\right) \tag{4}$$

where $\tilde{m}_T$ is the average of $\tilde{m}(t)$ over the integration time $T$. Similarly, the relative variance of $\hat{c}_{TA}$ can be obtained as

$$\left(\frac{\delta \hat{c}_{TA}}{\hat{c}_{TA}}\right)^2 = \frac{2\tilde{\tau}_b}{T\tilde{m}_T} \tag{5}$$

where $\tilde{\tau}_b$ is the modified average bound time. Interestingly, the relative variance of $\hat{c}_{TA}$ can be lower than the BP limit (Figure (2b)) which is due to the additional (spurious) events arising from the noise. From this result, one may think that the addition of noise somehow improves the reliability of estimation. However, this is a misperception since the parameters $\tilde{k}_+$ and $\tilde{k}_-$ are not practically obtainable as it requires the knowledge of the true signal average $m_T$ (See Appendix2). Therefore, to estimate the concentration, we need to use the actual binding and unbinding rates, i.e., $k_+$ and $k_-$ which makes the estimation biased. We define the biased estimated concentration $\tilde{c}_{TA}$ as

$$\tilde{c}_{TA} = \frac{\widetilde{m}_T}{1 - \widetilde{m}_T}\left(\frac{k_-}{k_+}\right) \tag{6}$$

The relative deviation, in this case, becomes, (See Appendix 2)

$$\frac{\langle \tilde{c}_{TA} - c_0 \rangle^2}{c_0^2} = \frac{2\tilde{\tau}_b}{T}\left(\frac{\widetilde{m}_T(1 - m_T)}{m_T(1 - \widetilde{m}_T)}\right)^2 + \left(\frac{\tilde{c}_{TA}}{c_0} - 1\right)^2 \tag{7}$$

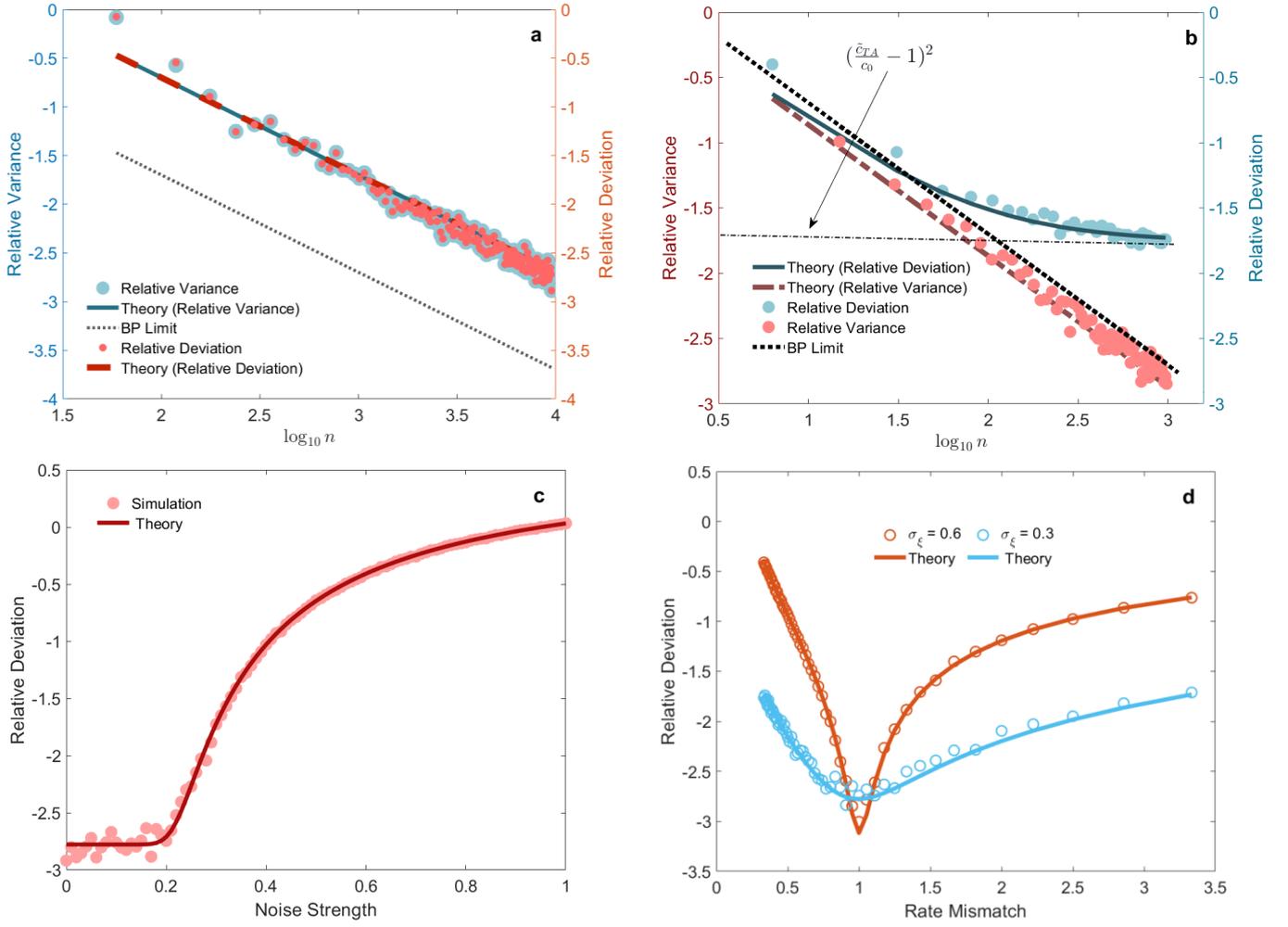

Figure 2. Relative variance $\left(\frac{\langle(\hat{c}-\langle\hat{c}\rangle)^2\rangle}{\langle\hat{c}\rangle^2}\right)$ and relative deviation $\left(\frac{\langle\hat{c}-c_0\rangle^2}{c_0^2}\right)$ in direct averaging (DA) and threshold-based averaging (TA) approaches: a. Comparison of relative variance and relative deviation in DA. As $\langle\hat{c}_{DA}\rangle = c_0$ in DA, the relative variance is equivalent to relative deviation. Both relative deviation and relative variance monotonically decrease with an increase in total time $T$. Due to the presence of measurement noise, the relative variance of DA is larger compared to the BP limit. [Parameters: $k_+ = k_- = 0.3, c_0 = 1, \sigma_\xi^2 = 1$]. b. Comparison of relative variance and relative deviation in TA: The relative variance in TA is lower compared to the BP limit

due to the additional events arising from the noise. Though relative variance monotonically decreases with an increase in total time $T$, relative deviation saturates at $\left(\frac{\tilde{c}_{TA}}{c_0} - 1\right)^2$. [Parameters: $k_+ = 0.3, k_- = 1, \sigma_\xi^2 = 0.3, c_0 = 1$] c. Relative deviation as a function of noise strength $(\sigma_\xi^2)$ for a given integration time $T$. [Parameters: $k_+ = 0.1, k_- = 0.03, c_0 = 1, T = 8000$] d. Relative deviation as a function of $k_+c/k_-$: For a given integration time $T$, when $k_+c = k_-$, relative deviation becomes equivalent to relative variance for TA and hence attains minima at $\frac{k_+c}{k_-} = 1$.[Parameters: $T = 16000, \sigma_\xi^2 = 0.3$]

We see that, while relative variance monotonically decreases with an increase in the integration time, the relative deviation saturates at $\left(\frac{\tilde{c}_{TA}}{c_0} - 1\right)^2$ (Figure 2b). For a fixed integration time, the relative deviation shows a transition between two regimes as shown in Fig. 2c. For low noise strength, the relative deviation coincides with the relative variance and remains close to the noise-free value of $2/n$ where $n$ is the total number of binding events [5]. Beyond some value of noise magnitude, the relative deviation starts to deviate significantly due to the appearance of spurious events due to noise and again becomes independent of noise in the high-noise limit as the probability of the signal being 1 or 0 approaches ½. The increase in the value of relative deviation arises due to the biased estimate of the ligand concentration, as mentioned before. Interestingly, this bias can be eliminated when the binding and unbinding rates are made equal i.e. $k_+c = k_-$ since this condition makes $\tilde{m}_T$ equal to $m_T$. Figure 2(d) demonstrates the effect of rate mismatch $(k_+c/k_-)$ on the relative deviation which attains a minimum at $\frac{k_+c}{k_-} = 1$.

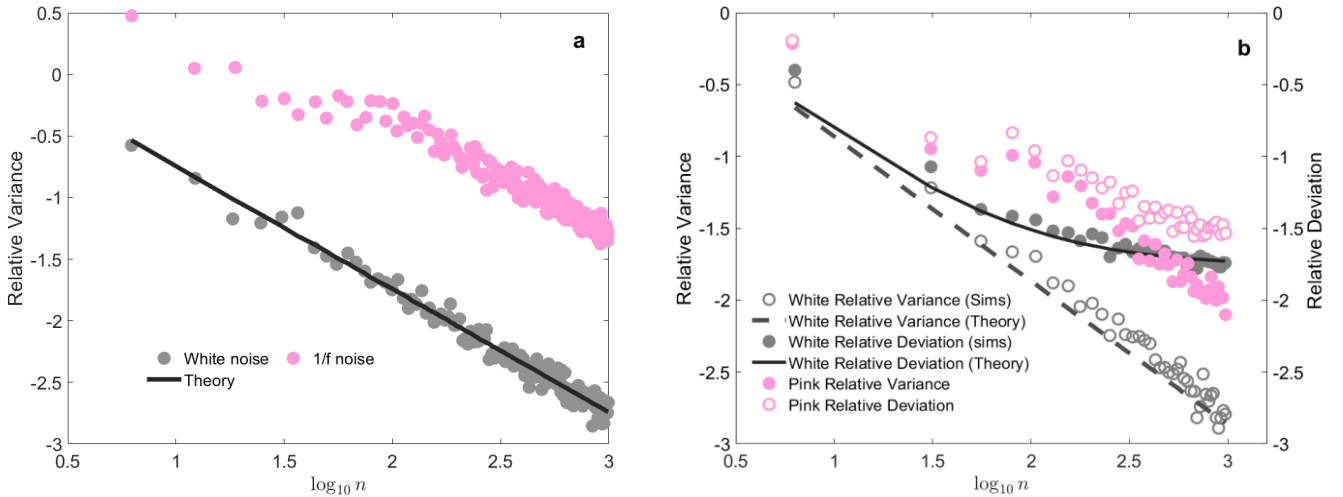

Figure 3. Effect of $1/f$ noise: a. Relative variance in TA b. Relative variance and relative deviation in DA.

We now consider the case of $1/f$ noise. Qualitatively, $1/f$ noise behaves identically as white noise. Quantitatively, for a given integration time $T$, the presence of $1/f$ noise has a more detrimental effect (Figure 3). Figure 3a shows the effect of $1/f$ noise on the relative variance in DA. This increase in relative variance can be understood as the reduction in the effective number of independent measurements due to correlations present in $1/f$ noise. In the case of TA, the relative deviation in the presence of $1/f$ noise also qualitatively behaves similar to the white noise, as shown in Figure 3b.

**Discussion**

In this work, we have rederived the Berg and Purcell's limit for a single receptor like system in the presence of additive zero-mean measurement noise. To this end, we considered two approaches. First, we directly averaged the measured signal which provides an accurate estimate of concentration since the noise is zero mean. However, the variance of the estimated concentration increases due to the noise, and hence the precision of estimation deteriorates. This effect is more prominent in case of colored noise such as $1/f$ noise compared to white noise. Second, we thresholded the measured signal to obtain a binary signal. Estimating concentration from the thresholded signal drastically improves the precision of estimation which can even surpass the zero noise Berg and Purcell limit. However, this improved precision comes at the cost of poor accuracy of estimation which unlike the precision doesn't improve with an increase in integration time. We presented our analysis in the context of artificial single nanopore sensors which although analogous to the single receptor system, possess some differences. One major difference is the binding rate constant $k_+$. For a single receptor system in the cellular context, $k_+$ only depends on the diffusion of the receptor molecules and the area of the receptor patch. However, in the case of the nanopore system, $k_+$ can be modulated using different techniques such as an external electric field [18], electrostatic traps [19], and bio-inspired fluid walls [20], etc. Therefore, we may be able to achieve the unbiased sensing regime (Figure 2d) for $k_+ c_0 = k_-$ by electric field tuning. Finally, although we analyze the situation of signal corrupted by independent additive noise, our analysis is equally applicable in situations where the signal is corrupted by other external interference such as the binding of spurious ligand molecules [21]. We expect our results to hold qualitatively in these situations also.

**Appendix 1 Derivation of Relative variance for DA**

In the DA approach, we estimate the concentration by directly averaging the measured signal ($m(t)$). If $m_T$ is the average of $m(t)$, the variance in $m_T$ can be written as

$$\langle m_T^2 \rangle = \frac{1}{T^2} \int_0^T dt' \int_0^T dt\, G(t-t') \tag{8}$$

where $G(\tau)$ is the autocorrelation of the signal $m(t)$ defined as

$$G(\tau) = \langle m(t)m(t+\tau) \rangle = \langle B(t)B(t+\tau) \rangle + \langle \xi(t)\xi(t+\tau) \rangle$$

If the integration time $T \gg \tau_b$, where $\tau_b$ is the average time interval for which a molecule stays bound to the receptor, then, following Berg and Purcell,

$$\langle m_T^2 \rangle - \langle m_T \rangle^2 = \frac{2}{T} \overline{B}(1-\overline{B})^2 \tau_b + \frac{1}{T^2} \int_0^T dt' \int_0^T dt\, \langle \xi(t)\xi(t') \rangle \tag{9}$$

Equation (9) represents the error in estimating $m_T$. Using propagation of errors, the relative variance of $\hat{c}_{DA}$ can be calculated as

$$\left( \frac{\delta \hat{c}_{DA}}{c_0} \right)^2 = \frac{2\tau_b}{Tm_T} + \frac{1}{T^2 m_T(1-m_T)^2} \int_0^T dt' \int_0^T dt\, \langle \xi(t)\xi(t') \rangle \tag{10}$$

which is Equation (3) in the main text. For white noise, Equation (10) reduces to

$$\left(\frac{\delta \hat{c}_{DA}}{c_0}\right)^2 = \frac{1}{Tm_T}\left(2\tau_b + \frac{\sigma_\xi^2}{m_T(1-m_T)^2}\right) \quad (11)$$

**Appendix 2 Derivation of relative variance and relative deviation in TA**

In the TA approach, we threshold the measured signal $m(t)$ before estimating concentration. If we choose $\mu$ as the threshold, for white noise, the mean of the thresholded signal $\tilde{m}(t)$ becomes

$$\tilde{m}_T = m_T\left(1 - 2Q\left(\frac{\mu}{\sigma}\right)\right) + Q\left(\frac{\mu}{\sigma}\right) \quad (12)$$

where $Q\left(\frac{\mu}{\sigma}\right)$ is the $Q$ function defined as $Q(x) = \frac{1}{\sqrt{2\pi}}\int_x^\infty \exp\left(-\frac{u^2}{2}\right)du$ and $m_T$ is the mean of the measured signal $m(t)$ before thresholding. As described in the main text, the concentration $\hat{c}_{TA}$ can be estimated from $\tilde{m}_T$ as

$$\hat{c}_{TA} = \frac{\tilde{m}_T}{1 - \tilde{m}_T}\left(\frac{\tilde{k}_-}{\tilde{k}_+}\right) \quad (13)$$

where $\tilde{k}_+$ and $\tilde{k}_-$ are the modified binding and unbinding rates. Since, $\tilde{m}(t)$ is a binary signal, under Markov condition, we can calculate the relative variance of the concentration $\hat{c}_{TA}$ estimated from $\tilde{m}(t)$ exactly as Berg and Purcell which becomes

$$\left(\frac{\delta \hat{c}_{TA}}{\hat{c}_{TA}}\right)^2 = \frac{2\tilde{\tau}_b}{T\tilde{m}_T} \quad (14)$$

where $\tilde{\tau}_b = \frac{\tilde{m}_T}{m_T\left(1-Q\left(\frac{\mu}{\sigma}\right)\right)\left(k_{off}+Q\left(\frac{\mu}{\sigma}\right)\right)+(1-m_T)Q\left(\frac{\mu}{\sigma}\right)\left(1-Q\left(\frac{\mu}{\sigma}\right)\right)}$. The estimation error in Equation 14 can be lower compared to the BP limit as $\frac{\tilde{\tau}_b}{\tilde{m}_T}$ can be smaller than $\frac{\tau_b}{m_T}$. However, to achieve this, one needs to obtain $\tilde{k}_+$ and $\tilde{k}_-$ from the signal $\tilde{m}(t)$. Since $\tilde{k}_- = 1/\tilde{\tau}_b$, to obtain $\tilde{k}_-$, one requires the knowledge of $m_T$ which is the average of the signal before thresholding. Therefore, $\tilde{k}_-$ is not practically obtainable from $\tilde{m}(t)$. Furthermore, it is not practically possible to obtain $\tilde{k}_+$ as the concentration $c$ is unknown. Therefore, one needs to use the true binding and unbinding rates i.e. $k_+$ and $k_-$ to estimate the concentration. This makes the estimated concentration $\tilde{c}_{TA} = \frac{\tilde{m}_T}{1-\tilde{m}_T}\left(\frac{k_-}{k_+}\right)$ different from the true concentration $c_0$, hence introducing bias. To address the issue of deviation of the estimate from the true concentration, we define the relative deviation as the variance of $\tilde{c}_{TA}$ around the true concentration $c_0$ as

$$\frac{\langle \tilde{c}_{TA} - c_0 \rangle^2}{c_0^2} = \frac{2\tilde{\tau}_b}{T}\left(\frac{\widetilde{m}_T(1-m_T)}{m_T(1-\widetilde{m}_T)}\right)^2 + \left(\frac{\tilde{c}_{TA}}{c_0} - 1\right)^2 \tag{15}$$

Interestingly, the effective bias in $\tilde{c}_{TA}$ can be nullified for $k_+ c = k_-$ by choosing the threshold $\mu = 0.5$ in Equation 12. For this condition, $\widetilde{m}_T$ becomes equal to $m_T$, since $\widetilde{m}_T = m_T = \frac{1}{2}$.